# High Harmonic Generation without Tunnel-Ionization

Jonathan Berkheim[1,2], Eliyahu Bordo[1,2], Eldar Ragonis[1,2], Lev Merensky[1,2] and Avner Fleischer[1,2]*

[1] *Raymond and Beverly Sackler Faculty of Exact Sciences, School of Chemistry, Tel Aviv University, Tel Aviv 6997801, Israel*
[2] *Tel-Aviv University center for Light-Matter-Interaction, Tel Aviv 6997801, Israel*
*Corresponding Author: avnerfleisch@tauex.tau.ac.il*

## Abstract

A new High Harmonic Generation (HHG) scheme, which doesn't rely on Tunnel-Ionization as the ionization mechanism but rather on Single-Photon Ionization, is theoretically proposed and numerically demonstrated. The scheme uses two driver fields: an extreme-ultraviolet driver which induces the ionization, and a circularly-polarized, co-rotating, two-color infrared driver carried at a fundamental frequency and its second harmonic which induces the recollision. Using Classical and time-dependent Schrödinger equation simulations of a model Argon atom, we show that in this scheme ionization is essentially decoupled from recollision. Releasing the HHG mechanism from being Tunneling-dependent reduces its degree of nonlinearity, which offers new capabilities in attosecond science, such as generation of High Harmonics from highly-charged ions, or from specific deep core electronic levels. It is shown that the emitted high harmonics involve the absorption of photons of one color of the infrared driver, and the emission of photons of the second color. This calls for future examination of the possible correlations between the emitted high harmonics.

High Harmonic Generation (HHG) is a central technique in attosecond science [1-3] and a prominent representative of extreme-nonlinear optical processes. In HHG intense infrared (IR) laser radiation is converted into the extreme ultraviolet (XUV) and soft x-ray regimes by interaction with gas, liquid [4-6] or solid medium [7-14]. Its strong nonlinearity stems from the event which initiates this 3-step process [15]: Tunnel-ionization (Tunneling) of a bound electron into the continuum. The two steps which follow are the propagation and recollision of the continuum electron with its remaining bound part, leading to the emission of short, attosecond bursts of radiation. This radiation source enables the tracking, exploration and control of ultrafast electronic processes in atoms, molecules [16-19] and solids [20] in a tabletop lab setup.

The nonlinear features of Tunneling, together with the inherent coupling between the ionization and recollision steps, impose fundamental and practical limitations in the HHG process. For instance, the maximal photon energy which can be achieved in the process, namely the cutoff energy $\hbar\Omega_{cut-off}$, is limited to roughly 90eV for Argon driven by 800nm linearly-polarized driver [21], irrespective of the driver intensity used. This results from the defocusing of the laser driver beam which accompanies excessive plasma generation. The plasma also adversely affects the coherent, phase-matched buildup of the HHG signal across the generation media [22]. Most importantly, high levels of ionization and ground-state depletion reduce the photorecombination cross section already at the single emitter level. Since Tunneling has extreme-nonlinear dependence on the IR driver's intensity, the restriction on the level of ionization limits the maximal effective driver intensity experienced by the media [23]. As the cutoff energy depends only linearly on the IR driver's intensity, $\hbar\Omega_{cut-off} = I_p + 3.17U_p$, (where $I_p$ is the ionization potential and $U_p = \frac{q^2 E_{IR}^2}{4m\omega^2}$ is the ponderomotive energy, with $q$, $m$ being the electron's charge and mass, and $E_{IR}$, $\omega$ being the driver's amplitude and frequency, respectively) the cutoff energy becomes limited as well. The nonlinear dependence of Tunneling on the ionization potential of the media $Ip$ imposes another limitation, as it excludes Tunneling from deep-core electronic levels over Tunneling from high-lying valence levels, or Tunneling from cationic species [24] when a mixture of neutral and ionized media is used [25].

Replacement of the Tunneling step in the HHG process with Single-Photon Ionization (SPI), e.g. by absorption of a high-energy XUV photon of frequency $\omega_{XUV}$, could in principle remove the abovementioned limitations. However when a linearly-polarized IR driver is used the recollision is adversely affected [26] by the initial momentum

$$k_0 = \sqrt{\frac{2m(\hbar\omega_{XUV} - Ip)}{\hbar^2}} \quad (1)$$

with which the photo-electron appears in the continuum according to the Photoelectric-effect ($\hbar$ is the Planck constant). Hence, in previous studies the inclusion of XUV drivers has been mostly utilized for the excitation of Rydberg states from which Tunneling by an IR field is easier [27-29], for the transfer electrons to the continuum for streaking or photoionization studies [30-31] or as a spectroscopic tool [32-33]. Only mild improvement of the HHG yield or cutoff energy has been obtained [32, 34-38].

Here we demonstrate a new HHG scheme in which SPI replaces Tunneling as the first step in the 3-step model, while still providing an intense recollision. It consists of shining the generation media by a combination of a high-frequency UXV driver and an IR driver. The large frequency difference between the drivers ensures that the evolution of the electron occurs on two different timescales. Hence, by virtue of a Born-Oppenheimer-like separation of variables, the two drivers take separate roles: the high-frequency XUV field efficiently liberates the electron to the continuum by SPI (1st step of the 3-step model) but hardly affects the propagation in the continuum other than inducing small quiver to the trajectory [39], while the IR driver is solely responsible for the propagation and recollision (2nd and 3rd steps of the 3-step model). In order to facilitate an intense recollision, the IR driver itself is a two-color, co-rotating circularly-polarized field. We analyze this scheme using both classical trajectory analysis and time-dependent Schrödinger equation (TDSE) simulations and show that the phase, polarization direction and frequency of the XUV driver determine the structure of the HHG spectra (HGS), while its intensity determine the HGS intensity.

In the context of HHG, a two-color circularly-polarized, co-rotating IR driver field with carrier frequencies of fundamental $\omega$ and its second harmonic ($r = 2$):

$$\mathbf{E}_{\mathbf{IR}}(t) = E_{IR}\left\{\left[-\cos\left(\omega t + \frac{\pi}{4}\right) - \cos\left(r\omega t + \frac{\pi}{4}\right)\right]\hat{x} + \left[\sin\left(\omega t + \frac{\pi}{4}\right) + \sin\left(r\omega t + \frac{\pi}{4}\right)\right]\hat{y}\right\} \quad (2)$$

is considered useless. For an electron initially at rest (as is the case when the electron is liberated by Tunneling), the field doesn't manage to return the electron back to the origin. This picture is consistent with photon angular momentum (spin) conservation considerations. As each driver photon in the Field of Eq. 2 already has a minimal spin of minus unity, the parametric conversion of these photons into a high-energy photon is not possible [40,41]. The Lissajous curve of this field is given by the blue trace of Fig. 1d. for $\omega = 0.02848$ (wavelength of $\lambda = 1600nm$), $r = 2$, and $E_{IR} = 0.03$ (intensity of $I = 6.32 \cdot 10^{13} \left[ W/cm^2 \right]$ for each color)]. We now show that this field could in fact generate efficient HHG provided that an XUV field which liberates the electron to the continuum with a large initial velocity is added. We examine the classical trajectories resulting from the propagation of a free electron of charge $q$ and mass $m$ (atomic units are used throughout, $m = -q = 1$), initially at the origin, freed at time $t_i$ into the two-color IR field of Eq.2. Integrating the classical Newton second-law equation twice, from the release time $t_i$, trajectories which return to the origin (termed "closed" or "recolliding" trajectories) at time $t_r$ exist if the initial velocity $\mathbf{v} = \begin{bmatrix} v_x \\ v_y \end{bmatrix}$ of the electron fulfills:

$$0 = (t_r - t_i)\mathbf{v} + \frac{q}{m}\int_{t_i}^{t_r} dt' \int_{t_i}^{t'} dt'' \mathbf{E}_{IR}(t'') \quad (3)$$

For each value of the initial velocity $\mathbf{v}$, a group of several closed trajectories might exist, each having a unique release time and kinetic energy upon recollision $E_{k,i}(\mathbf{v}) = \frac{1}{2m}\left[ m\mathbf{v} + q\int_{t_i}^{t_r} dt' \mathbf{E}_{IR}(t') \right]^2$. Fig. 1a shows the average kinetic energy upon recollision $\bar{E}_k(\mathbf{v}) = \frac{\sum_i W_i \cdot E_{k,i}(\mathbf{v})}{\sum_i W_i}$ of all trajectories having the initial velocity $\mathbf{v}$, in units of the Ponderomotive potential of the two-color IR field $U_p = \frac{q^2 E_{IR}^2}{2m\omega^2}\left(1 + \frac{1}{r^2}\right)$. The contribution of each trajectory is weighted according to its recombination cross section, which is assumed to inversely depend on the recolliding electronic wavepacket lateral spreading only, according to $W_i = (t_r - t_i)^{-2}$. Similarly, average values of the release times $\bar{t}_i(\mathbf{v}) = \frac{\sum_i W_i \cdot t_i(\mathbf{v})}{\sum_i W_i}$ and recollision times $\bar{t}_r(\mathbf{v}) = \frac{\sum_i W_i \cdot t_r(\mathbf{v})}{\sum_i W_i}$ are shown in units of the optical period $T = \frac{2\pi}{\omega}$ in Fig1b,c respectively. It is seen that the return of the electron to the origin is possible only if large values of initial velocities $\mathbf{v}$ are given to the electron. In particular, when $\mathbf{v} = \begin{bmatrix} -0.125 \\ -0.875 \end{bmatrix} \equiv \mathbf{v}^{cut\,off}$ (red dot in Fig.1a) the cutoff trajectory and the maximal average kinetic energy upon recollision $\bar{E}_k(\mathbf{v}^{cut\,off}) \approx 2.246 U_p$ are obtained, with release and recollision times being $\bar{t}_i(\mathbf{v}^{cut\,off}) = 0.469T$ and $\bar{t}_r(\mathbf{v}^{cut\,off}) = 1.077T$. A specific cutoff trajectory representing this group of cutoff trajectories, with values $\mathbf{v} = \begin{bmatrix} -0.125 \\ -0.875 \end{bmatrix}$, $t_i = 0.469T$, $t_r = 1.067T$, $E_k(t_r = 1.067T) \approx 2.276 U_p$ is given by the red trace in Fig. 1d. This trajectory starts at (returns to) the origin when the IR field's instantaneous amplitude is close to zero (maximal), in contrast to the cutoff trajectory in the traditional HHG scheme (based on Tunnel-ionization induced by the IR field).

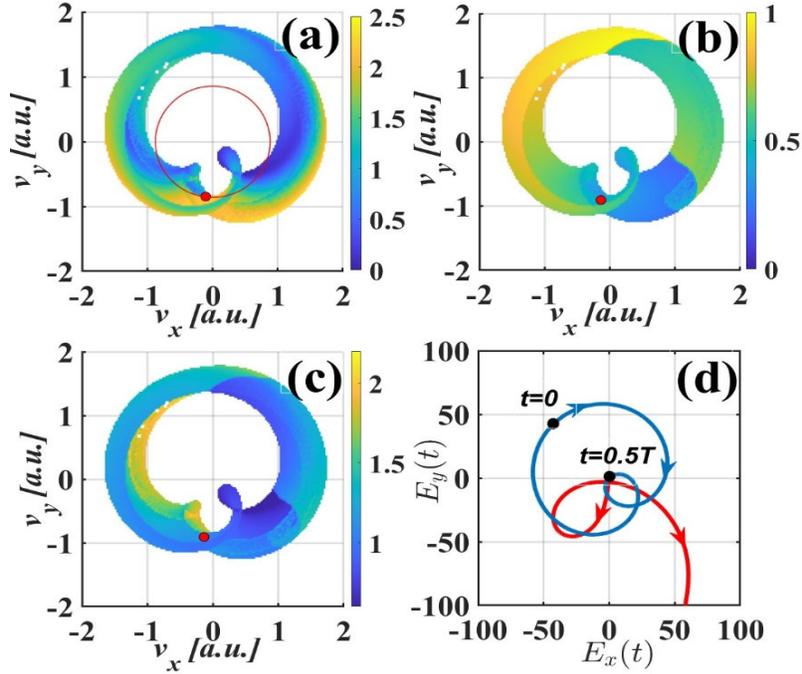

FIG 1 (color online) (a): Averaged kinetic energy upon recollision $\bar{E}_k(\mathbf{v})$ of all trajectories with initial velocity $\mathbf{v} = \begin{bmatrix} v_x \\ v_y \end{bmatrix}$ in units of the IR driver ponderomotive energy $U_p$. (b) $\bar{t}_i(\mathbf{v})$ [T]. (c) $\bar{t}_r(\mathbf{v})$ [T]. (d) Lissajous curve (scaled × 1000) of the field in Eq.2. with $r = 2$ (blue trace) and an exemplary cutoff trajectory (red trace). The red circle in (a) has a radius of 0.8839. Along this circle, the average kinetic energy is peaked for the initial velocity $\mathbf{v} = \begin{bmatrix} -0.125 \\ -0.875 \end{bmatrix}$.

SPI by high-frequency XUV driver field is capable of liberating electrons into the continuum with large initial velocity- a task which can't be accomplished when the ionization is carried out by Tunneling. Then, recollision could be induced by the field of Eq.2. We demonstrate this by numerically solving the TDSE for a single electron in a 3D model of Argon atom. The TDSE is solved in the length gauge ($\hbar = 1$),

$$\left[ -\frac{\hbar^2}{2m}\nabla^2 + V(r) - q\mathbf{r}\cdot\mathbf{E}(t) \right]\Psi(\mathbf{r},t) = i\hbar\frac{\partial}{\partial t}\Psi(\mathbf{r},t) \qquad (4)$$

and the field-free potential is taken to be $V(r) = \frac{-Z(r)}{\sqrt{r^2 + a}}$ ($r = \sqrt{\mathbf{r} \cdot \mathbf{r}}$) with $Z(r) = 1 + 0.2719 e^{-0.25r}$ and $a = 0.09192$, which yields the correct two lowest bound states of Argon atom [42]. The driver electric field is composed of two parts $\mathbf{E}(t) = f_{IR}(t) \mathbf{E}_{\mathbf{IR}}(t) + f_{XUV}(t) \mathbf{E}_{\mathbf{XUV}}(t)$, where the XUV field

$$\mathbf{E}_{\mathbf{XUV}}(t) = E_{XUV} \left[ -\cos(\omega_{XUV} t) \hat{x} + \sin(\omega_{XUV} t) \hat{y} \right] \quad (5)$$

is circularly polarized with same helicity as the other two colors and $E_{XUV}$ being its amplitude. Here both drivers envelopes $f_{IR}(t), f_{XUV}(t)$ are trapezoid with 7 cycles rising and falling edges, and 6 cycles plateau (in units of $T$). We slightly detune ($r = 1.95$) the second harmonic of the IR driver which causes the Lissajous curve to slowly drift throughout the pulse. For times $10T < t < 11T$ it acquires the trace given in the inset of Fig. 2. The $\mathbf{E}_{\mathbf{XUV}}(t)$ field amplitude is $E_{XUV} = 0.01$ ($I = 7.02 \cdot 10^{12} \left[ \frac{W}{cm^2} \right]$), and its carrier frequency is usually $\omega_{XUV} = 34\omega$ unless otherwise stated. The HGS were obtained by Fourier-transforming the time dependent dipole acceleration expectation value $\mathbf{a}(t) = \langle \Psi(\mathbf{r},t) | -\nabla V | \Psi(\mathbf{r},t) \rangle$.

Figure 2 shows the TDSE simulation results. A reference spectra where the system is driven by the two-color co-rotating IR field alone (no XUV field) is shown in the yellow curve. As expected, besides few low-order harmonics, no HHG is generated. The IR driver induces ionization by Tunneling, with the ionization probability at the end of the pulse being $P_{ion}(t = 20T) \approx 0.0327$. Supplementing the IR driver field with the XUV driver changes the picture dramatically. Several traces are shown (see also Fig S1 for more traces): $\omega_{XUV} = 26\omega$ (green curve, $P_{ion}(t = 20T) \approx 0.273$), $\omega_{XUV} = 34\omega$ (red, $P_{ion}(t = 20T) \approx 0.121$), $\omega_{XUV} = 42\omega$ (blue, $P_{ion}(t = 20T) \approx 0.071$) and $\omega_{XUV} = 50\omega$ (cyan, $P_{ion}(t = 20T) \approx 0.0513$). In all cases intense HHG is obtained, comprised of discrete peaks, a plateau and a cut-off. This seemingly contradiction to spin conservation (all driver photons are circularly-polarized with same helicity) will be resolved later. Out of the traces shown, the HGS in the cutoff region ($\Omega_{cut\,off} \approx 75\omega$) is maximized for $\omega_{XUV} = 34\omega$. This is not surprising in light of the classical trajectory results presented in Fig.1. With that XUV frequency, the cutoff recolliding trajectories are efficiently generated since the correct initial momentum is given to the electron by the XUV field: $k_0 = \sqrt{\frac{2m(\omega_{XUV} - Ip)}{\hbar^2}} = \sqrt{\frac{2m(34\omega - Ip)}{\hbar^2}} = 0.8821$, in accordance with the magnitude of the trajectory's initial velocity $|\mathbf{v}^{cut\,off}| = \sqrt{(-0.125)^2 + (-0.875)^2} = 0.8839$. Also, with $\bar{E}_k(\mathbf{v}^{cut\,off}) \approx 2.246 U_p \approx 54.7\omega$ and $Ip \approx 0.579 \approx 20.34\omega$, the group of cutoff trajectories are generating the cut-off HHG emission around $\Omega_{cut\,off} \approx 76\omega$, as observed. For comparison, we show that the HGS of our new scheme is comparable in intensity to the HGS obtained with the Tunneling-based scheme of a two-color, counter-rotating IR driver (black curve, $P_{ion}(t=20T) \approx 0.0402$) [40,43-46].

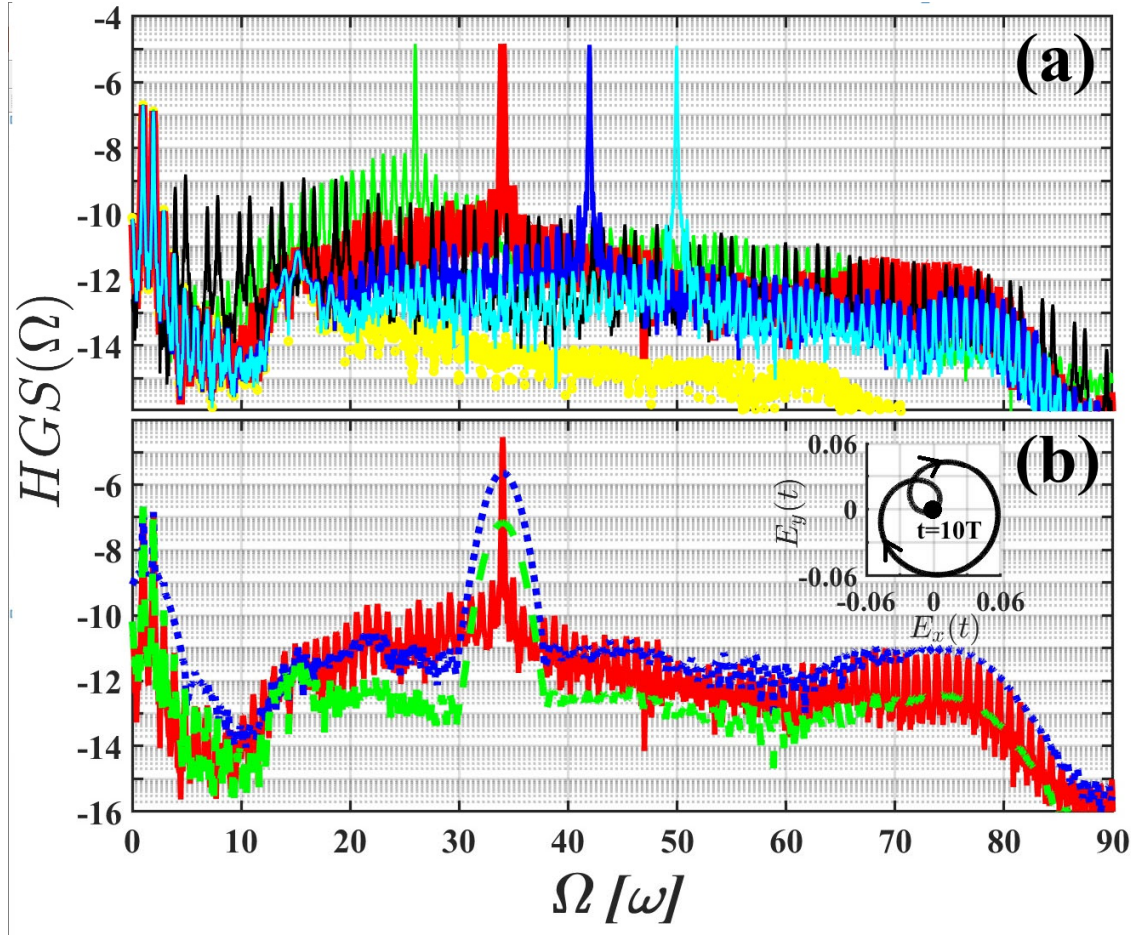

FIG 2 (color online) (a): HGS (log scale) for the co-rotating ($r=1.95$) IR+XUV driver for $\omega_{XUV}=26\omega$ (green curve), $\omega_{XUV}=34\omega$ (red), $\omega_{XUV}=42\omega$ (blue) and $\omega_{XUV}=50\omega$ (cyan); Reference spectra where the system is driven by the co-rotating IR field only (no XUV field, yellow curve) and when the system is driven by a two-color, counter-rotating IR field only (no XUV field, black curve) are also shown. (b): HGS (log scale) for the co-rotating ($r=1.95$) trapezoid IR driver and short Gaussian pulse of circularly-polarized XUV driver with $\omega_{XUV}=34\omega$ and $E_{XUV}=0.02108$ (dashed green curve), and a 6-times larger XUV amplitude (dotted blue curve). Red curve- same as in (a). (Inset): Lissajous curve (scaled x 1000) of the IR driver (black), which passes the origin at the arrival time of the XUV Gaussian pulse $t_{XUV}=10T$. The Lissajous curve is rotated by π with respect to the one shown in Fig.1d due to the drift imposed by the frequency ratio ($r=1.95$) of the IR driver constituents;

In SPI, the ionization yield $P_{ion}$ has linear dependence on the intensity of the XUV driver, $P_{ion} \propto |E_{XUV}|^2$ and quadratic dependence $P_{ion} = \langle \psi_{cont.} | \psi_{cont.} \rangle$ on the part of the continuum wavefunction $\psi_{cont.}$. Hence, the existence of recollision implies that the HGS intensity, which is related to the absolute-squared continuum-to-bound transition amplitude, should be linear in the XUV driver's intensity as well, $HGS(\Omega) \sim |\langle \psi_b | -\nabla V | \psi_{cont.} \rangle|^2 \sim o(|\psi_{cont.}|^2) \sim |E_{XUV}|^2$. These two predictions are verified in Fig.2b which shows the HGS simulated with all parameters as before ($\omega_{XUV} = 34\omega$), but with the trapezoid XUV envelope replaced with a short Gaussian pulse $f_{XUV}(t) = \exp\left[-2\ln 2 \left(\frac{t - t_{XUV}}{\tau_{XUV}}\right)^2\right]$ whose duration is $\tau_{XUV} = 0.25T$. In order to generate the cutoff trajectories, the short XUV pulse arrives at time $t_{XUV} = 10T$, when the IR field ($r = 1.95$) crosses the origin. This time is very close to the release time $\bar{t}_i(\mathbf{v}^{cut\,off}) = 9.969T$ predicted by the classical trajectory calculations. The green curve shows the results with the XUV pulse with amplitude $E_{XUV} = 0.02108$ ($P_{ion}(t = 20T) \approx 0.03798$) and the blue curve shows the results where the XUV amplitude has been increased 6-fold to $E_{XUV} = 0.1265$ ($P_{ion}(t = 20T) \approx 0.2055$). Subtracting the Tunneling-induced ionization ($P_{ion}(t = 20T) \approx 0.0327$) from the total ionization probabilities gives the contribution of the XUV ionization event alone; a ratio of (0.2055-0.0327)/(0.03798-0.0327)~32.7 is obtained, in good agreement with the predicted ratio of 36. The two HGS traces also show a 36-fold increase in the HGS in the spectral region of the plateau and cutoff, as anticipated. Moreover, the continuous-shape of the HGS (rather than a comb-like structure) with THE short-pulse XUV ionizer is another indication that recollision occurs in the process. As the electron is born during a single short time window, only a single recollision event occurs. These results prove that our proposed HHG scheme indeed involves recollision and SPI-initiated ionization, and that they are separated.

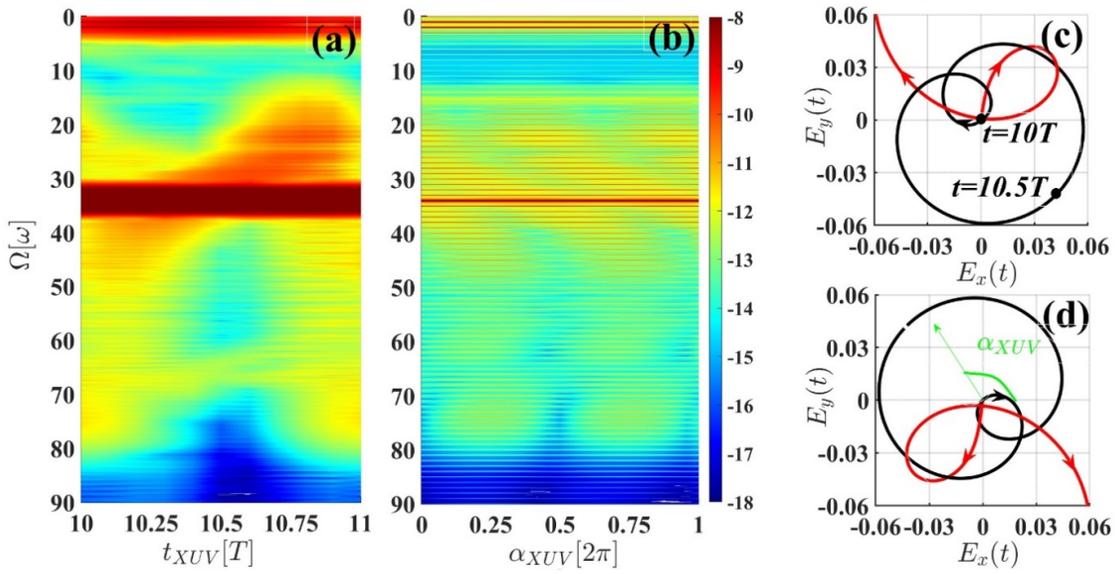

FIG 3 (color online) (a): HGS (log scale) for the co-rotating trapezoid IR driver ($r = 1.95$) and short Gaussian pulse of circularly-polarized XUV driver with $\omega_{XUV} = 34\omega$ for various arrival times $t_{xuv}$ of the XUV driver field. (b): HGS for the co-rotating trapezoid IR driver ($r = 2$) and linearly-polarized XUV driver for various polarization directions of the XUV driver field. (c)- Lissajous curve of the IR driver (black) and cutoff trajectory with $\mathbf{v} = \begin{bmatrix} 0.125 \\ 0.875 \end{bmatrix}$ (red) corresponding to the simulations of panel (a) with $t_{xuv} = 10T$. (d)- Lissajous curve of the IR driver (black) and cutoff trajectory with $\mathbf{v} = \begin{bmatrix} -0.125 \\ -0.875 \end{bmatrix}$ (red) corresponding to the simulations of panel (b) with XUV pulses polarized along the y-direction ($\alpha_{XUV} = \pm\frac{\pi}{2}$).

As our scheme suggests, a temporal synchronization between the XUV ionization and the IR driver is critical in order to ensure the recollision. This was indeed verified (Fig. 3a) by analyzing the dependence of the HGS on $t_{XUV}$. As anticipated, the HGS in the cutoff region is maximized for $t_{XUV} \approx 10T$ since then the Lissajous curve of the two-color IR field crosses the origin (Fig. 3c) Moreover, a directional synchronization is needed as well. This is demonstrated by fixing the Lissajous curve for the IR driver in space ($r = 2$) and taking a linearly-polarized XUV field whose polarization direction creates an angle $\alpha_{xuv}$ with the x-axis. Our initial state has an 1s spherical symmetry, which results in a Malus-type sin-square dependence of the HGS yield (in the plateau and cutoff) on $\alpha_{xuv}$ (Fig. 3b). This is in accordance with the anticipated linear dependence of the HGS on the flux of XUV-generated photoelectrons, the trajectory picture, and the p-shaped ionization distribution of electrons in the continuum typical of SPI from a state with s-symmetry. The maximal HGS yield is obtained when the XUV polarization direction coincides with the y-axis, since this leads to the birth of recollision trajectories (Fig. 3d). On the other hand, if the XUV polarization direction coincides with the x-axis, photoelectrons are released in all directions but the y-direction, leading to a great suppression of the HGS.

Finally, we relate the semiclassical description described above to the Photonic-picture description of HHG, and Photonic conservation laws. In the simulations presented above (excluding those of Fig.3b) the spins of all 3 colors of driver photons have been deliberately chosen as minus unity $\sigma_1 = \sigma_2 = \sigma_{XUV} = -1$. In such driver configuration spin conservation excludes the absorption of driver photons and emission of a **single** high-energy photon, since it is not possible for a photon spin to exceed unity. Since nevertheless intense HHG is observed, this means that more than a single photon is emitted in the HHG process. Fig. S2 shown the HGS and ellipticity-helicty product of a simulation with $r = 1.95$, $\omega_{XUV} = 34\omega$ and longer trapezoid driver pulses. The narrow harmonic peaks obtained enable the unique identification of harmonic emission channels according to their energy and polarization state. As it has been unequivocally identified that the HHG process is linear in the intensity of the XUV driver, only a single XUV driver photon is annihilated. All in all, four families of emission channels have been identified, all yielding circularly-polarized photons. The first family are emission channels of the form $\Omega_{(-n,n,1)}$ (where the integer numbers in the parenthesis refer to the number of annihilated photons from each color $\omega_1$, $\omega_2$, $\omega_{XUV}$ of the drivers, respectively, and a negative integer describes an emission of driver photons back to the driver field). This channel has an energy $\Omega_{(-n,n,1)} = (-n) \cdot \omega_1 + n \cdot \omega_2 + 1 \cdot \omega_{XUV} = n(\omega_2 - \omega_1) + \omega_{XUV}$ and spin $\sigma_{(-n,n,1)} = (-n) \cdot (-1) + n \cdot (-1) + 1 \cdot (-1) = -1$. A specific channel which belongs to the first family is for instance $\Omega_{(-20,20,1)} = 53\omega$, $\sigma_{(-20,20,1)} = -1$. In this channel, 20 photons of the second IR driver and a single photon of the XUV driver are absorbed (a total of 21 photons) and 21 photons are emitted: 20 photons back to the first color driver, and a single high-energy photon. The re-emission of driver photons removes excess spin from the system and enables this emission channel. A representative of the second family is e.g., $\Omega_{(-24,22,1)} = 52.9\omega$, $\sigma_{(-24,22,1)} = +1$. In the other two families photons are emitted back to the second IR color, e.g. $\Omega_{(20,-20,1)} = 15\omega$, $\sigma_{(20,-20,1)} = -1$ and $\Omega_{(16,-18,1)} = 14.9\omega$, $\sigma_{(16,-18,1)} = +1$. In all channels IR photons of one color are annihilated and IR photons of the second color are emitted. This fact, in particular in light of the structure of the emission channels $\Omega_{(-n,n,1)}$ and $\Omega_{(n,-n,1)}$, suggests that the high-harmonic photons might be correlated. This should be checked in future studies.

To summarize, we have presented a new HHG scheme in which the electron is liberated into the continuum by SPI rather than by Tunneling. The recollision is induced by a well-synchronized two-color, co-rotating IR field with correct frequencies and amplitudes to return the electron to the ion despite of the initial velocity it acquired from the XUV ionizing field. Our scheme has several unique properties: first, the ionization step is separated from the propagation/recollision steps. The ionization is performed by the XUV field, while the propagation and recombination are controlled by the IR field. This means that the shape of the HGS could be controlled separately from the HHG efficiency: the IR field controls the shape and cutoff of the spectrum (by virtue of the recollisions and attochirp curve), while the spectrum's intensity depends linearly on the intensity of the XUV driver. Second, the ionization by SPI allows the generation of High Harmonics from electronic levels which are hardly accessible when Tunneling is concerned, e.g. HHG from charged ions, or from specific deep core electronic levels. Third, in order to generate recolliding trajectories, the XUV frequency and the IR frequency and intensity must be related according to Eqs. 1,3. In addition, the temporal synchronization and relative polarization-directions between the XUV and IR drivers determines which trajectories participate in the recollision. By combining bright XUV sources, as found in Free-Electron-Laser facilities with the IR laser as in our scheme, efficient attosecond pulse sources could be generated at those facilities. To the best of our knowledge, this is the first HHG mechanism which relies on the emission of several photons and not just a single one. It is reasonable to assume that some or all of the emitted harmonics are probably correlated.

This work was supported by the Israel Science Foundation (Grant No. 524/19).

## Supplemental Material

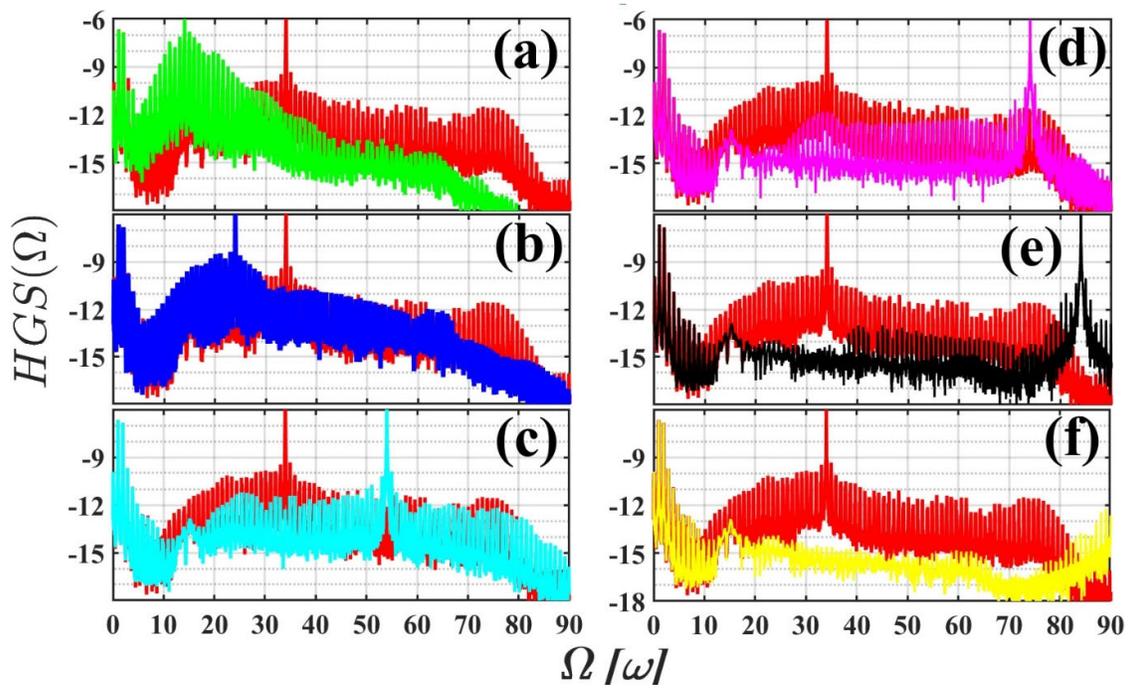

FIG S1 (color online) (a)- HGS for the co-rotating IR+XUV driver with different frequencies of the XUV driver field: $\omega_{XUV} = 14\omega$ (green curve), $\omega_{XUV} = 34\omega$ (red); (b)- $\omega_{XUV} = 24\omega$ (blue). (c)- $\omega_{XUV} = 54\omega$ (cyan). (d)- $\omega_{XUV} = 74\omega$ (magenta). (e)- $\omega_{XUV} = 84\omega$ (black). (f)- $\omega_{XUV} = 94\omega$ (yellow).

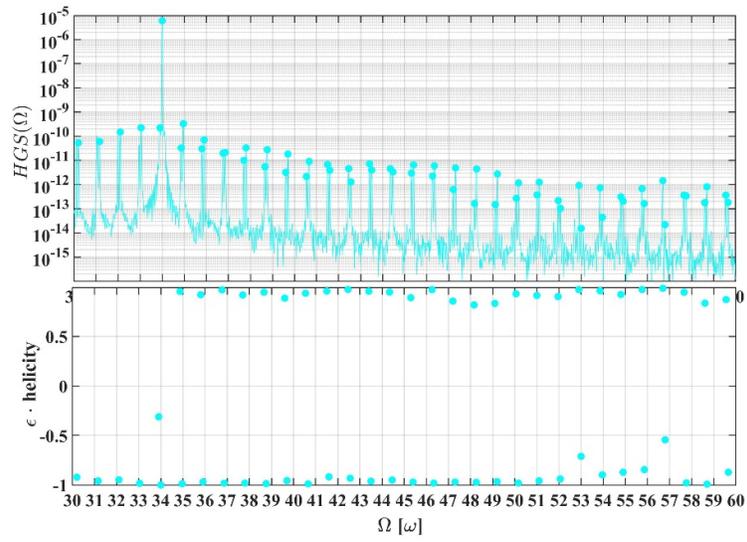

FIG S2 (color online) (a) HGS for the co-rotating ($r = 1.95$) IR+XUV driver $\omega_{XUV} = 34\omega$ and long trapezoid pulses. (b) ellipticity-helicity of the obtained harmonics.